\documentclass[12pt,preprint2,apj]{emulateapj}

\shorttitle{Search for Cross-Correlations of UHECR with BL Lacs}
\shortauthors{The HiRes Collaboration}

\begin{document}

\title{Search for Cross-Correlations of Ultra--High-Energy Cosmic Rays 
with BL Lacertae Objects}
\author{
R.U.~Abbasi,\altaffilmark{1}
T.~Abu-Zayyad,\altaffilmark{1}
J.F.~Amann,\altaffilmark{2}
G.~Archbold,\altaffilmark{1}
K.~Belov,\altaffilmark{1}
J.W.~Belz,\altaffilmark{3}
S.~BenZvi,\altaffilmark{4}
D.R.~Bergman,\altaffilmark{5}
S.A.~Blake,\altaffilmark{1}
J.H.~Boyer,\altaffilmark{4}
G.W.~Burt,\altaffilmark{1}
Z.~Cao,\altaffilmark{1}
B.M.~Connolly,\altaffilmark{4}
W.~Deng,\altaffilmark{1}
Y.~Fedorova,\altaffilmark{1}
J.~Findlay,\altaffilmark{1}
C.B.~Finley,\altaffilmark{4}
W.F.~Hanlon,\altaffilmark{1}
C.M.~Hoffman,\altaffilmark{2}
M.H.~Holzscheiter,\altaffilmark{2}
G.A.~Hughes,\altaffilmark{5}
P.~H\"{u}ntemeyer,\altaffilmark{1}
C.C.H.~Jui,\altaffilmark{1}
K.~Kim,\altaffilmark{1}
M.A.~Kirn,\altaffilmark{3}
B.C.~Knapp,\altaffilmark{4}
E.C.~Loh,\altaffilmark{1}
M.M.~Maestas,\altaffilmark{1}
N.~Manago,\altaffilmark{6}
E.J.~Mannel,\altaffilmark{4}
L.J.~Marek,\altaffilmark{2}
K.~Martens,\altaffilmark{1}
J.A.J.~Matthews,\altaffilmark{7}
J.N.~Matthews,\altaffilmark{1}
A.~O'Neill,\altaffilmark{4}
C.A.~Painter,\altaffilmark{2}
L.~Perera,\altaffilmark{5}
K.~Reil,\altaffilmark{1}
R.~Riehle,\altaffilmark{1}
M.D.~Roberts,\altaffilmark{7}
D.~Rodriguez,\altaffilmark{1}
M.~Sasaki,\altaffilmark{6}
S.R.~Schnetzer,\altaffilmark{5}
M.~Seman,\altaffilmark{4}
G.~Sinnis,\altaffilmark{2}
J.D.~Smith,\altaffilmark{1}
R.~Snow,\altaffilmark{1}
P.~Sokolsky,\altaffilmark{1}
R.W.~Springer,\altaffilmark{1}
B.T.~Stokes,\altaffilmark{1}
J.R.~Thomas,\altaffilmark{1}
S.B.~Thomas,\altaffilmark{1}
G.B.~Thomson,\altaffilmark{5}
D.~Tupa,\altaffilmark{2}
S.~Westerhoff,\altaffilmark{4}
L.R.~Wiencke,\altaffilmark{1}
and A.~Zech\altaffilmark{5}\\ 
(The HiRes Collaboration)
}

%---------------------

\altaffiltext{1}{University of Utah,
Department of Physics and High Energy Astrophysics Institute,
Salt Lake City, UT 84112.}

\altaffiltext{2}{Los Alamos National Laboratory,
Los Alamos, NM 87545.}

\altaffiltext{3}{University of Montana, Department of Physics and Astronomy,
Missoula, MT 59812.}

\altaffiltext{4}{Columbia University, Department of Physics and
Nevis Laboratories, New York, NY 10027: finley@physics.columbia.edu,
westerhoff@nevis.columbia.edu.}

\altaffiltext{5}{Rutgers --- The State University of New Jersey,
Department of Physics and Astronomy, Piscataway, NJ 08854.}

\altaffiltext{6}{University of Tokyo,
Institute for Cosmic Ray Research,
Kashiwa City, Chiba 277-8582, Japan.}

\altaffiltext{7}{University of New Mexico,
Department of Physics and Astronomy,
Albuquerque, NM 87131.}

\begin{abstract}
Data taken in stereo mode by the High Resolution Fly's Eye (HiRes)
air fluorescence experiment are analyzed to search for correlations between
the arrival directions
of ultra--high-energy cosmic rays with the positions of BL Lacertae objects.
Several previous claims of significant correlations between BL Lacs
and cosmic rays observed by other experiments are tested.  These
claims are not supported by the HiRes data.  
However, we verify a recent analysis
of correlations between HiRes events and a subset of confirmed BL Lacs 
from the 10th Veron Catalog, and we study this
correlation in detail.  Due to the {\it a posteriori} nature of the search, 
the significance level cannot be reliably estimated and the correlation 
must be tested independently before any claim can be made.  
We identify the precise 
hypotheses that will be tested with statistically independent data.
\end{abstract}


\section{Introduction}

\setcounter{footnote}{0} % emulateapj needs footnotes reset after authorlist

Among the most striking astrophysical phenomena today is the
existence of cosmic ray particles with energies up to and exceeding
$10^{20}$\,eV.  It is currently unknown where and how these
particles are accelerated to such energies and how they
travel astronomical distances without substantial energy loss.

In an attempt to understand the origin of these particles, the limited world 
data set of ultra--high-energy cosmic ray arrival directions has been 
subjected to extensive searches for correlations with the positions 
of objects from known astrophysical source classes.  
In particular, significant correlations between 
subsets of BL Lacertae objects and cosmic ray arrival directions 
recorded by the Akeno Giant Air Shower Array (AGASA) and 
the Yakutsk experiment have been claimed 
\citep{Tinyakov:2001nr,Tinyakov:2001ir,Gorbunov:2002hk}.
Searches for correlations with BL Lac objects are well-motivated.
BL Lacs are a subclass of blazars, which are active galaxies in which the jet 
axis happens to point almost directly along the line of sight.
The EGRET instrument on board the Compton Gamma Ray Observatory (CGRO) has
firmly established blazars as sources of high energy $\gamma$-rays
above 100\,MeV \citep{Hartman:1999fc}, and several BL Lac objects
have been observed at TeV energies with ground-based air Cherenkov
telescopes \citep[see][for a summary]{Horan:2003wg}.  
High energy $\gamma$-rays could be
by-products of electromagnetic cascades from energy
losses associated with the acceleration of ultra--high-energy cosmic
rays and their propagation in intergalactic space \citep{Berezinskii:1990,
Coppi:1996ze}.

However, the claimed correlations between ultra--high-energy cosmic 
ray arrival directions and BL Lac objects are controversial.  
A problematic aspect of the claims is the procedure used to establish 
correlations and evaluate their statistical significance.  Several 
authors \citep{Tinyakov:2001nr,Tinyakov:2001ir,Gorbunov:2002hk} 
explicitly tuned 
their selection criteria to assemble catalogs that show a maximum 
correlation with arrival directions of cosmic rays above some energy.  
An unbiased chance probability for these correlations can then only be 
arrived at if the claim is tested on a statistically independent data 
set.  Since the available data set is small, this rigorous procedure 
is often abandoned, and instead an attempt is made to estimate 
a statistical penalty factor to compensate for the number of trials 
involved in the tuning.  ``Hidden'' trials, unfortunately, make these
{\it a posteriori} estimates
highly unreliable, and claims of BL Lac correlations
have been criticized on these grounds \citep{Evans:2002ry,Stern:2005fh}.  
Additionally, in some cases it has been shown that 
statistically independent data sets do not 
confirm the correlations \citep{Torres:2003ee}.  

The operation of the stereoscopic High Resolution Fly's Eye (HiRes)
air fluorescence detector is providing a large data set of cosmic ray
events with unprecedented angular resolution for the study of small-scale
anisotropy and source correlations.  In this paper, we report on searches
for correlations between BL Lac objects and HiRes stereoscopic events 
observed between 1999 December and 2004 January.
The quality cuts applied to this data sample
are described in detail in \citet{apjl2004,apj2005}.

The outline of the paper is as follows: in Section \ref{sec:method},
we describe our use of the maximum likelihood method for correlation
searches with multiple sources.  In Section \ref{sec:tests}, we consider
the previous claims of correlations between BL Lacs and ultra--high-energy
cosmic rays and test them using HiRes stereo data.
In Section \ref{sec:newcorr}, we perform a general search for correlations
between the HiRes data set and confirmed BL Lacs. In Section \ref{sec:summary},
we summarize the results and discuss further studies.

\section{Maximum Likelihood Method}\label{sec:method}

\subsection{Description}

To address a number of shortcomings of binned analyses of cosmic ray
arrival directions, we have
recently applied an {\it unbinned} maximum likelihood method in 
the search for point sources of ultra--high-energy cosmic rays \citep{apj2005}.
This approach
uses the probability density function for each individual event rather than 
requiring a fixed bin size.  Two important advantages 
of this method are the ability to accommodate events with different errors, 
and to give weighted sensitivity to angular separations---avoiding
the loss of information that follows from 
choosing an angular separation cut-off.
With minor modifications, the same maximum likelihood method 
can also be used to search for correlations 
with a specified list of potential sources.

The premise involved in the maximum likelihood analysis is that the data sample
of $N$ events consists of $n_s$ source events which came from 
some source position(s)
in the sky, and $N-n_s$ background events.  A background event arrives  
according to the probability distribution given by the detector
exposure to the sky, $R({\bf x})$,
where ${\bf x}$ are equatorial coordinates.  The true arrival direction
of a source event is the location of the source ${\bf s}$, but the event is
observed somewhere near ${\bf s}$ according to the probability distribution
$Q_i({\bf x},{\bf s})$, where $Q_i$ depends on the angular uncertainty
in arrival direction of the $i$th event.

Because it is not known whether a given event is a source or background event,
the probability distribution function (or ``partial'' probability) 
for the $i$th event is a weighted sum
of the source and background probability distributions:
\begin{equation}\label{eq:partial}
P_i({\bf x}) = \frac{n_{s}}{N}\,Q_{i}({\bf x},{\bf s})
                + \frac{N-n_{s}}{N}\,R({\bf x})~~.
\end{equation}

This describes the distribution of arrival directions
under a single-source hypothesis.  For a 
hypothesis with $M$ sources, we must modify $Q$ to include
multiple source locations.  We will assume for this analysis 
that the sources have equal luminosity.  In this case, we only need to 
compensate for the varying exposure of the detector 
to different parts of the sky: the probability for a 
source event to come from the $j$th source is proportional to the detector
exposure $R({\bf s}_j)$ to the source location ${\bf s}_j$.  
The total source probability distribution $Q_i^{tot}$ for the $i$th event 
is then the weighted sum of the individual source probabilities:
\begin{equation}
Q_{i}^{tot}({\bf x}) = \sum_{j=1}^{M} R({\bf s}_j) Q_i({\bf x},{\bf s}_j)
             /\sum_{k=1}^{M} R({\bf s}_k)~~.
\end{equation}

Replacing $Q_i$ in Eq.~\ref{eq:partial} with $Q_{i}^{tot}$,
we evaluate the partial probability of the $i$th event at its observed location
${\bf x}_i$:
\begin{equation}
P_i({\bf x}_i) = \frac{n_{s}}{N}\,Q_{i}^{tot}({\bf x}_i)
                + \frac{N-n_{s}}{N}\,R({\bf x}_i)~~.
\end{equation}

The likelihood for the entire set of $N$ events is then given by:
\begin{equation}
{\mathcal L}(n_{s}) = \prod_{i=1}^{N} P_{i}({\bf x}_{i})~~.
\end{equation}
The best estimate for the number of events contributed by  
the sources is determined by finding the value of 
$n_{s}$ that maximizes $\mathcal{L}$.

Because the value of the likelihood function depends on the
number of events, a more useful quantity than $\mathcal{L}$
is the likelihood ratio $\mathcal{R}$:
\begin{equation}
{\mathcal R}(n_{s}) = \frac{{\mathcal L}(n_{s})}
        {{\mathcal L}(0)}\nonumber
    = \prod_{i=1}^{N}~
        \left\{\frac{n_{s}}{N}
           \left(\frac{Q_{i}^{tot}({\bf x}_{i})}
                      {R({\bf x}_{i})}
           -1\right)
        +1\right\}
\end{equation}
where ${\mathcal L}(0)$ is the likelihood function of the 
{\it null hypothesis} 
($n_{s}=0$).  In practice, we maximize $\ln\mathcal R$, which is equivalent to 
maximizing $\mathcal{L}$.  

The significance of the resulting $\ln\mathcal R$ can be estimated using
$\chi^2 = 2\ln\mathcal R$.  When $n_s$ is positive, this 
agrees well with the $\chi^2$ distribution for one degree
of freedom.  Because $n_s$ corresponds to the {\it excess}
number of events correlating with source positions, a negative
best-fit value for $n_s$ will occur whenever there are fewer events
near source positions than expected.  Negative $n_s$ values are 
not physically meaningful in the point-source search, 
but they are useful for evaluating significances.
To distinguish an excess in correlations from a deficit, we 
assign the negative solution $\chi = - \sqrt{2 \ln\mathcal R}$ 
when the best-fit $n_s$ is negative.  We can then check the significance
estimated from the $\chi^2$ distribution
by performing the same likelihood analysis on simulated data sets
and ranking them according to their $\chi$ values.
We will use $\mathcal F$ to 
denote the fraction of simulated, isotropic event sets which yield a
value of $\chi$ greater than or equal to that of the data.

\begin{deluxetable*}{lccccccccccc}
\tabletypesize{\scriptsize}
\tablecaption{\label{tab:bintest}Previously claimed correlations between
BL Lacs and cosmic rays, and tests with independent HiRes data}
\tablewidth{0pt}
\tablehead{ & Sample   &           & 
     CR Data and Energy   & \#  & \multicolumn{3}{c}{Binned Analysis} & &
     \multicolumn{3}{c}{Max. Like. Analysis} \\
\cline{6-8} \cline{10-12}
\\[-6pt]
 & (\# Objects) & Reference & 
     Threshold (EeV)      & Ev. & Bin Size & Pairs & Prob. & &
     $\ln\mathcal R$  & $n_s$ & $\mathcal F$}
\startdata
 Claim 1: & A (22) & TT01 &
     AGASA$>$48, Yak.$>$24 & 65 &
     $2.5^{\circ}$ & 8 & $<10^{-4}$ \\
 Test: & & &
     HiRes $>24$ & 66 &
     $2.5^{\circ}$ & 0 & 1.00 & &
     (0) & (0) & 0.75 \\[6pt]
 Claim 2: & B (157) & TT02 &
     AGASA $>40$ & 57 &
     $2.5^{\circ}$ & 12 & 0.02 \\
 Test: & & &
     HiRes $>40$ & 27 &
     $2.5^{\circ}$ & 2 & 0.78 & &
     (0) & (0) & 0.26 \\[6pt]
 Claim 3: & C (14) & G02 &
     AGASA$>$48, Yak.$>$24 & 65 &
     $2.9^{\circ}$ & 8 & $10^{-4}$ \\
 Test: & & &
     HiRes $>24$ & 66 &
     $2.9^{\circ}$ & 1 & 0.70 & &
     (0) & (0) & 0.68 \\[-6pt]
\enddata
\tablerefs{[TT01] \citet{Tinyakov:2001nr}; [TT02] \citet{Tinyakov:2001ir};
[G02] \citet{Gorbunov:2002hk}; [G04] \citet{Gorbunov:2004bs}.
In [TT02] and [G02], the authors also attempt to 
correct for the deflections of 
charged primaries by the galactic magnetic field; these results are not
considered here.}
\end{deluxetable*}

\subsection{Implementation}

For the source probability function
$Q_{i}$ we employ a circular Gaussian of width $\sigma_i$ corresponding to the
angular uncertainty of the $i$th event, as estimated by the stereo event
reconstruction.  The mean of the angular uncertainty of the HiRes
stereo events is slightly larger at
lower energies, growing from $\left<\sigma \right> = 0.44^{\circ}$ above 
$10^{18}$\,eV to $\left<\sigma \right> = 0.60^{\circ}$ below $10^{17.75}$\,eV.
This of course is accounted for by the use of individual errors in the
maximum likelihood analysis.

Because the geometrical acceptance of the detector is a function of energy,
we determine the background probability distribution $R({\bf x})$ 
in two different ways.
For large event samples ($\gtrsim 1000$), we generate a background distribution
from a full time-swapping of the data itself: the equatorial coordinates of 
each event are recalculated using all of the recorded event times, and 
$R({\bf x})$ is
the cumulative map of all of these virtual event locations convolved with a
circular Gaussian function for smoothing.  For small event samples (e.g. the
271 events above $10^{19}$\,eV) the data set is too sparse to generate
a useful time-swapped map.  
Instead, we rely on a full detector simulation to estimate
the local geometrical acceptance, and convolve this with the event times
to generate $R({\bf x})$.  
This procedure and the detector simulation are described
in more detail in \citet{apjl2004}.

\section{Tests of Previous Correlations Observed with AGASA and Yakutsk Data}
\label{sec:tests}

We briefly review some past claims of correlations between cosmic ray
arrival directions and BL Lacs,
and then test these with HiRes data under the same conditions.
All samples of BL Lacs are selected from objects classified as
``BL'' in Table 2 of the Veron Catalog of Quasars and AGN,
9th or 10th Editions \citep{Veron:9th,Veron:10th}.  

\begin{itemize}

\item
{\it Sample A:} described in \citet{Tinyakov:2001nr}, 
contains 22 BL Lacs from the Veron 9th Catalog selected on the basis of
optical magnitude ($m<18$), redshift ($z>0.1$ or unknown), 
and 6\,cm radio flux ($F_{6}>0.17$\,Jy).

\item
{\it Sample B:} described in \citet{Tinyakov:2001ir}, 
contains 157 BL Lacs from the Veron 10th Catalog with optical magnitude
$m<18$.

\item
{\it Sample C:} described in \citet{Gorbunov:2002hk}, 
consists of 14 BL Lacs from the 
Veron 10th catalog that were selected by the authors on the basis
of possible association with identified and unidentified
$\gamma$-ray sources in the Third EGRET Catalog \citep{Hartman:1999fc}.

\end{itemize}

Table~\ref{tab:bintest} shows the correlations originally claimed using
these BL Lac samples and cosmic ray data from the AGASA and Yakutsk 
experiments.  The energy thresholds and angular bin sizes vary from
analysis to analysis as shown.   
The results of testing each claim as nearly as possible
with an equivalent set of HiRes data are also presented.
Both a binned analysis with the originally used bin size
and a maximum likelihood analysis using the point spread function of
individual HiRes events are performed.  In the binned analysis,
the number of event-object pairs with angular separation less than the
bin size are counted, and the probability for the same or greater number
of pairs is evaluated using simulated isotropic event sets.
None of the three previous claims of correlations based on other 
cosmic ray data sets are confirmed by the tests.  Each test,
in fact, finds a deficit or no excess of HiRes events correlating with BL Lacs,
indicated by (0) values for $\ln \mathcal R$ and $n_s$.  The fraction 
$\mathcal F$ of simulated sets with stronger correlation than the data
is calculated as described above.

In the tests of Claims 1 and 3, 
the size of the HiRes event sample is comparable to the 
size of the combined AGASA and Yakutsk event samples.  
Assuming a Poisson distribution
with mean number of event - BL Lac pairs given by the AGASA and
Yakutsk results, the observed
number of HiRes - BL Lac pairs excludes the claimed correlations
at a confidence level greater than 99\,\% in each case.  In the test 
of Claim 2, the HiRes event sample is smaller than that of AGASA.  
Here the 
claimed correlation is excluded at the 90\,\% confidence level.

\section{Recent Correlations Observed with HiRes Data}
\label{sec:newcorr}

Recently, the published HiRes events above $10^{19}$\,eV
were analyzed by \citet{Gorbunov:2004bs}, 
and correlations with the BL Lacs of Sample
B were claimed at the $10^{-3}$ level.  The analysis used a fixed
bin size of $0.8^{\circ}$, which the authors argued is optimal for 
a point-source search given the HiRes angular resolution.
We verify this analysis by applying the maximum likelihood method
to the same data set and source sample,
and find $\ln\mathcal R = 6.08$ for $n_s = 8.0$; 
the fraction of Monte Carlo sets with higher 
$\ln\mathcal R$ is $\mathcal F = 2 \times 10^{-4}$.

\citet{Gorbunov:2004bs} analyzed the entire set of HiRes 
events above $10^{19}$\,eV because the individual event energies were
not published.  Therefore this energy threshold was not tuned to 
maximize correlations with BL Lacs.  However, because the original claim
\citep{Tinyakov:2001ir}
was based on AGASA data with energies above $4\times 10^{19}$\,eV, the
correlation in \citet{Gorbunov:2004bs} 
does not confirm a previous claim, but rather represents
a new hypothesis.  This is demonstrated by the fact that the 
HiRes data shows no excess correlation with
this sample of BL Lacs when the same $4\times 10^{19}$\,eV 
energy threshold is used, 
as indicated in the test of Claim 2 in Table~\ref{tab:bintest}.

The observed correlation warrants further investigation.  
We report on extending the analysis to lower energy HiRes data 
and to the rest of the confirmed BL Lacs in the Veron catalog.

\subsection{Event Sample: Energy Dependence of Correlations}

\begin{figure}
\epsscale{1.2} 
\plotone{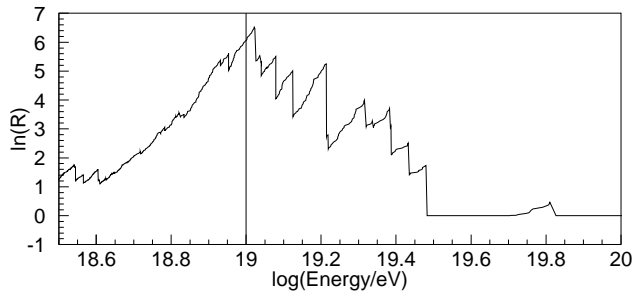}
\caption{\label{fig:escan}$\ln\mathcal R$ result as a function of minimum
energy threshold of the HiRes data set.  The $10^{19}$\,eV energy threshold
of the published data is indicated.}
\end{figure}

An important question is whether and how the observed correlation
depends on the energy threshold.  Figure~\ref{fig:escan} shows
the result of the same maximum likelihood analysis
above, performed repeatedly using increasing energy thresholds from 
$10^{18.5}$\,eV to $10^{20}$\,eV.  The $10^{19}$\,eV
threshold corresponding to the published data set is indicated, 
and it clearly stands out as the threshold that gives a local maximum
in the significance of the correlation.  

One of the motivations for using an energy threshold in small-scale
anisotropy searches 
is that charged cosmic ray primaries are subject to deflection
by galactic and extragalactic magnetic fields.  
The highest energy primaries will be least deflected, and consequently 
will be the best candidates for correlation studies.  
However, an {\it a priori} energy 
threshold cannot be established, because detailed knowledge of the galactic
and extragalactic magnetic fields is lacking.

Figure~\ref{fig:escan} indicates 
that most of the correlation comes from events with energies between 
$10^{19}$\,eV and $10^{19.5}$\,eV.  
At these energies, it is generally assumed that the Galactic 
magnetic field will deflect a proton primary by many degrees; nuclei will
be deflected even more.  In spite of this, 
the correlations are consistent with the $\sim 0.5^{\circ}$ scale 
of the detector angular resolution.  
This might imply that the correlated primary cosmic rays are neutral, 
thus removing the motivation for restricting the 
analysis to events
with energies above some arbitrary threshold. A search for
correlations with the entire HiRes stereo data set of 4495 events at all
energies is justified.

Applying the analysis to the entire data set and Sample B gives
$\ln\mathcal R = 6.16$ for $n_s = 31$, with $\mathcal F = 2 \times 10^{-4}$.
This of course includes the effect of the correlated
events above $10^{19}$\,eV;
for the independent sample of 4224 events below $10^{19}$\,eV, we find
$\ln\mathcal R = 3.10$ for $n_s = 22$, with $\mathcal F = 6\times 10^{-3}$.

\subsection{Source Sample}

The Sample B of BL Lacs discussed above consists of 157 confirmed BL Lacs 
in the 10th Veron Catalog with optical magnitude $m<18$ and which are classified 
as ``BL''.  The rest of the confirmed BL Lacs are classified as ``HP'' 
(high polarization).  It is natural to ask about these objects
as well.  Indeed, of the six blazars which have confirmed detections in 
$\gamma$-rays at TeV energies, half are classified as ``HP'' and half as 
``BL'' in the Veron catalog.

We apply the same cut on optical magnitude $m<18$ as in Sample B
to the ``HP'' objects,
and arrive at a sample of 47 objects.
The result of the maximum likelihood analysis applied to this 
independent sample of BL Lacs and the HiRes events
above $10^{19}$\,eV is $\ln\mathcal R = 3.13$ for $n_s = 3.0$,
with $\mathcal F = 6 \times 10^{-3}$.  
We also perform the same analysis on the events below
$10^{19}$\,eV.  No excess is found.

\begin{deluxetable}{ccc}
\tablewidth{0pt}
\tablecaption{\label{tab:hiresbl}HiRes --- BL Lac Correlation Results: 
Fraction $\mathcal F$ of simulated HiRes sets
with stronger correlation signal.}
\tablehead{Source Sample (\# Obj.) & All Energies   & $E>10$\,EeV}
\startdata
``BL'' (157) & $2\times 10^{-4}$  & $2\times 10^{-4}$ \\
``HP'' (47)  & 0.3                & $6\times 10^{-3}$ \\
``BL''+``HP'' (204) & $5\times 10^{-4}$ & $10^{-5}$ \\[-6pt]
\enddata
\tablecomments{Correlations are with confirmed BL Lacs in Table 2 of the 
Veron 10th Catalog \citep{Veron:10th}, classified as either ``BL'' or ``HP,''
with $m<18$.}
\end{deluxetable}

The results for HiRes events of energy above $10^{19}$\,eV and all energies
with BL Lacs classified as ``BL,'' ``HP,'' or  ``BL'' and ``HP'' combined are 
summarized in Table~\ref{tab:hiresbl}.

The equivalent analyses have been performed 
on the same classes of BL Lacs with $m\ge 18$:
no excess correlation is found in any of these cases.  
It is apparent from these results that the $m < 18$
cut which was identified in \citet{Tinyakov:2001ir} 
as optimal for AGASA also isolates the BL Lac objects which show
excess correlations with HiRes events.
Under the BL Lac source hypothesis, of course,
it is not unreasonable to expect the closer and more 
luminous objects to contribute more strongly.  However,
since the Veron catalog is not a uniform sample of BL Lac objects,
the interpretation of this cut may involve a more complicated interplay of
selection effects from the underlying surveys which make up the 
catalog.\footnote{
The Veron catalog strives to be ``complete'' 
only in the sense of a complete survey of the literature and catalog 
of all known BL Lacs; it does not represent an unbiased statistical 
sample of BL Lacs in any way \citep{Veron:9th,Veron:10th}.  
This does not exclude the possibility of using subsets 
of the catalog to identify correlations with cosmic rays, 
but it means that any inferences about the BL Lacs based on such 
correlations may be highly biased and simply an artifact of the 
underlying combination of different surveys in the catalog.}

\subsection{TeV Blazars}

Among the closest and brightest of the ``BL'' and ``HP'' BL Lacs 
are six which are confirmed sources of TeV $\gamma$-rays \citep{Horan:2003wg}. 
Five of these, shown in Table~\ref{tab:hirestev}, 
are high in the northern sky and well within the field of view of HiRes.  
We perform the maximum likelihood analysis on this set of objects
using all of the HiRes data, and find $\ln\mathcal R = 4.78$ for
$n_s = 5.6$ with $\mathcal F = 10^{-3}$.  
For just the HiRes events above $10^{19}$\,eV,
the result is $\ln\mathcal R=6.15$ for $n_s=2.0$ with 
$\mathcal F=2\times 10^{-4}$.
In Table~\ref{tab:hirestev}, 
we show the results for treating each blazar in turn as 
a single source hypothesis.

\begin{deluxetable}{ccccccc}
\tablewidth{0pt}
\tablecaption{\label{tab:hirestev}TeV Blazar Correlation Results with HiRes
events (all energies)}
\tablehead{\multicolumn{3}{c}{TeV Blazars}    &  &\multicolumn{3}{c}{Correlation Results}\\
\cline{1-3}\cline{5-7}\\[-6pt]
Name        & z     & V Mag & & $n_s$ & $\ln\mathcal R$ & $\mathcal F$}
\startdata
Mrk 421     & 0.03  & 12.9  & & 0.3 & 0.04 &  0.2 \\
H1426+428   & 0.13  & 16.5  & & (0)\tablenotemark{a} & (0)  &  0.4 \\
Mrk 501     & 0.03  & 13.8  & & 3.3 & 5.27 &  $6\times 10^{-4}$ \\
1ES1959+650 & 0.05  & 12.8  & & 2.0 & 2.87 &  $8\times 10^{-3}$ \\
1ES2344+514 & 0.04  & 15.5  & & (0)\tablenotemark{b} & (0)  &  0.7 \\
\hline
Combined Set &      &       & & 5.6                & 4.78 & $10^{-3}$ \\[-6pt]
\enddata
\tablenotetext{a}{No excess: $n_s<3.5$ at 90\% confidence level.}
\tablenotetext{b}{No excess: $n_s<2.4$ at 90\% confidence level.}
\end{deluxetable}

\section{Results and Discussion}
\label{sec:summary}

In this paper, we have used an unbinned maximum likelihood method to analyze
correlations of ultra--high-energy cosmic ray arrival directions with 
BL Lac objects in the Veron 10th Catalog.  We first tested previous claims
of correlations between BL Lacs and cosmic rays which were based on AGASA
and Yakutsk data.  Using the independent HiRes stereo data set,  
these correlation claims are excluded 
at the 99\,\% (Claims 1 and 3 in Table~\ref{tab:bintest}) or 
90\,\% confidence level (Claim 2).

\begin{deluxetable}{ccc}
\tablewidth{0pt}
\tablecaption{\label{tab:summary}HiRes --- BL Lac Correlation Summary: Fraction $\mathcal F$ of simulated HiRes sets
with stronger correlation signal.}
\tablehead{Source Sample (\# Obj.)         & All Energies       & $E>10$\,EeV}
\startdata
``BL'' Objects, $m<18$ (157)    & $2\times 10^{-4}$  & $2\times 10^{-4}$ \\
Confirmed BL Lacs, $m<18$ (204) & $5\times 10^{-4}$  & $10^{-5}$ \\
Confirmed TeV Blazars (6)       & $10^{-3}$          & $2\times 10^{-4}$ \\[-6pt]
\enddata
\tablecomments{All samples are contained within Table 2 of the
Veron 10th Catalog.  The samples overlap and are {\it not} independent:
``Confirmed BL Lacs'' combines ``BL'' and ``HP'' classified BL Lacs;
TeV Blazars are a subset of the confirmed BL Lacs.}
\end{deluxetable}

However, we have verified the observation by~\citet{Gorbunov:2004bs}
that the set of HiRes stereo events with energies above
$10^{19}$\,eV shows an excess of events correlated with confirmed BL Lacs 
marked as ``BL'' in the Veron 10th Catalog.  We emphasize that the 
observed correlation does not confirm a previous claim, because it  
requires a lower energy threshold.  It can only be confirmed with new data.

We have explored the extension of the analysis to 1) HiRes events of all 
energies, and 2) the rest of the confirmed BL Lacs (labeled ``HP'') in
the Veron 10th Catalog.  In each case, correlations at the significance level
of $\sim 0.5\%$ are found.  While statistically independent from the
above result, these are not strictly tests of that claim.
However, the combination offers well-defined hypotheses
which can be tested with new data.

The results of combining the analysis of 
low and high energy events and ``BL''
and ``HP'' BL Lacs are summarized in Table~\ref{tab:summary}.  Also shown
are the results for HiRes events and the subset
of BL Lacs which are confirmed sources of TeV $\gamma$-rays.

The analyses described here have only been performed on the data recorded
through 2004 January.  
The HiRes detector will continue observations through the end of 
2006 March.  By that time the independent sample of data since
2004 January is expected to reach 
approximately 70\,\% of the size of the sample analyzed here.  
This will provide an 
opportunity to test the correlations in Table~\ref{tab:summary}.  
We note that while the correlation signals appear stronger for the 
events above $10^{19}$\,eV, 
a conservative approach which includes consideration of the entire 
data set will avoid the possibility that a real correlation has been 
``over-tuned'' by an arbitrary threshold and is missed in a future analysis.

As mentioned earlier, real correlations on the scale of the detector
angular resolution would suggest neutral cosmic ray primaries for these events,
or at least that the primaries were neutral 
during significant portions of their journey through galactic and
extragalactic magnetic fields.  Primaries such
as neutrons and photons are problematic, however, because of short
mean free paths ($\sim$ a few Mpc) at these energies.  
As a fluorescence detector, HiRes can study the height of the maximum 
of the shower development, a parameter that has some sensitivity 
to the primary particle type.  Showers induced by photons, for example, tend to
develop lower in the atmosphere than those induced by nucleons.  
An analysis is underway to address the question whether the correlated HiRes       
stereo events are compatible with gamma-induced showers.
\\

\acknowledgments
The HiRes project is supported by the National Science Foundation under
contract numbers NSF-PHY-9321949, NSF-PHY-9322298, NSF-PHY-9974537,
NSF-PHY-0098826, NSF-PHY-0245428, by the Department
of Energy Grant FG03-92ER40732, and by the Australian Research Council.
The cooperation of Colonels E. Fisher and G. Harter, the US Army and
Dugway Proving Ground staff is appreciated.
We thank the authors of CORSIKA for providing us with
the simulation code.

\bibliography{ms}

\end{document}